\documentclass[journal=jacsat,manuscript=article]{achemso}

\usepackage[version=3]{mhchem} 
\usepackage[utf8]{inputenc} 
\usepackage[T1]{fontenc}    
\usepackage{hyperref}       
\usepackage{url}            
\usepackage{booktabs}       
\usepackage{amsfonts}       
\usepackage{amsmath}
\usepackage{nicefrac}       
\usepackage{microtype}      
\usepackage{array} 
\usepackage{varwidth}
\usepackage{graphicx}
\usepackage{multirow}
\usepackage{subfig}
\usepackage{amssymb}
\usepackage{array}
\usepackage{longtable}

\usepackage[nomarkers,figuresonly]{endfloat}

\usepackage{siunitx}
\usepackage{math_macro}

\newcolumntype{C}[1]{>{\centering\arraybackslash}p{#1}}
\usepackage[flushleft]{threeparttable}

\author{Amr H. Mahmoud}
\affiliation{Department of Medicinal Chemistry and Molecular Pharmacology, College of Pharmacy, Purdue University, 575 Stadium Mall Drive, West Lafayette, Indiana 47906, United States}
\alsoaffiliation{Department of Pharmaceutical Sciences, University of Basel, Klingelbergstrasse 50, 4056 Basel, Switzerland}

\author{Jonas F. Lill}
\affiliation{Department of Pharmaceutical Sciences, University of Basel, Klingelbergstrasse 50, 4056 Basel, Switzerland}

\author{Markus A. Lill}
\affiliation{Department of Pharmaceutical Sciences, University of Basel, Klingelbergstrasse 50, 4056 Basel, Switzerland}
\email{markus.lill@unibas.ch}
\phone{++41 61 2076135}

\title[An \textsf{achemso} demo]
  {Graph-convolution neural network-based flexible docking utilizing coarse-grained distance matrix}

\abbreviations{Graph-convolution neural network, distance matrix, flexible docking, protein-ligand interactions}
\keywords{American Chemical Society, \LaTeX}

\begin{document}

\begin{abstract}
Prediction of protein-ligand complexes for flexible proteins remains still a challenging problem in computational structural biology and drug design. Here we present two novel deep neural network approaches with significant improvement in efficiency and accuracy of binding mode prediction on a large and diverse set of protein systems compared to standard docking. Whereas the first graph convolutional network is used for re-ranking poses the second approach aims to generate and rank poses independent of standard docking approaches. This novel approach relies on the prediction of distance matrices between ligand atoms and protein C$_\alpha$ atoms thus incorporating side-chain flexibility implicitly.

\end{abstract}

\section{Introduction}

Structure-based drug design is an essential tool and an important pillar in Computer-aided Drug Design (CADD) for efficient lead discovery and optimization. 
CADD methods such as docking aim to identify novel binders to a target protein and to predict the structure of protein-ligand complexes.
Docking is still widely applied using a rigid protein as template in CADD projects, ignoring the representation of the different conformations that the binding-site can assume. 


The importance of modelling protein flexibility in docking, however, has been recognized already in early docking studies. In 1999 Murray et al.\cite{murray1999sensitivity} demonstrated the shortcomings of docking to rigid proteins by carrying out rigid cross docking for three enzymes (thrombin, thermolysin and the influenza virus neuroaminidase). 
Each ligand was docked to the protein structures of all complexes available for each enzyme. The authors found that in 51\% of the cases the program failed to dock the small molecules directly and highlighted the importance of modeling protein flexibility in computational docking\cite{murray1999sensitivity}. Later, Englebienne \& Moitessier showed that the accuracy of many scoring functions can be deteriorated by protein flexibility and solvation\cite{englebienne2009docking}. 
 A large number of reviews discuss the importance of incorporating protein flexibility in docking algorithms while focusing on side-chain, backbone and domain movements necessary for the protein to accommodate different ligands.
 \cite{Lill2011,totrov2008flexible,durrant2010computer,henzler2010pursuit,a2011accounting}.


Incorporating protein flexibility into molecular docking is a difficult optimization problem involving a large number of degrees of freedom that represent the receptor flexibility. 
Approaches to incorporate receptor flexibility range from the use of soft-core potentials, multiple protein structures (ensemble docking) or the active sampling of protein conformations during energy optimization of the ligand  (induced-fit docking)\cite{ferrari2004soft,sherman2006novel}. 
Due to the computational complexity of the problem, for many practical uses, flexible docking is still a challenging task and full incorporation of protein flexibility is computationally not feasible. 
In summary, there is a significant demand for efficient algorithms to handle flexible proteins in docking.

Several limitations for improving flexible docking methodology  exist. 
Most of the published methods are optimized or benchmarked on limited data sets with a very limited set of targets \cite{sherman2006novel}. While this practice was acceptable in the past due to the limitation of computational resources, this is no longer the case nowadays. Validation has to be carried out using large test sets with a wide range of different targets. Only such a validation procedure allows to identify the shortcomings of current and newly developed algorithms. This is essential for systematic improvement of flexible protein docking methodologies.

Additionally, flexible docking is more resource and time intensive than rigid docking. The application of flexible docking to virtual screening of large libraries is still unrealistic. The number of degrees of freedom in flexible proteinn docking is significantly higher compared to rigid docking, leading to an increase in rate of false-positives and intensive usage of computational resources. Thus, there is a crucial need to develop new methods that rely on efficient algorithms and heuristics to lessen the computational requirements and allow an accurate widespread implementation of flexible docking engines.

\begin{figure}
 \centering
 
 \includegraphics[width=\textwidth]{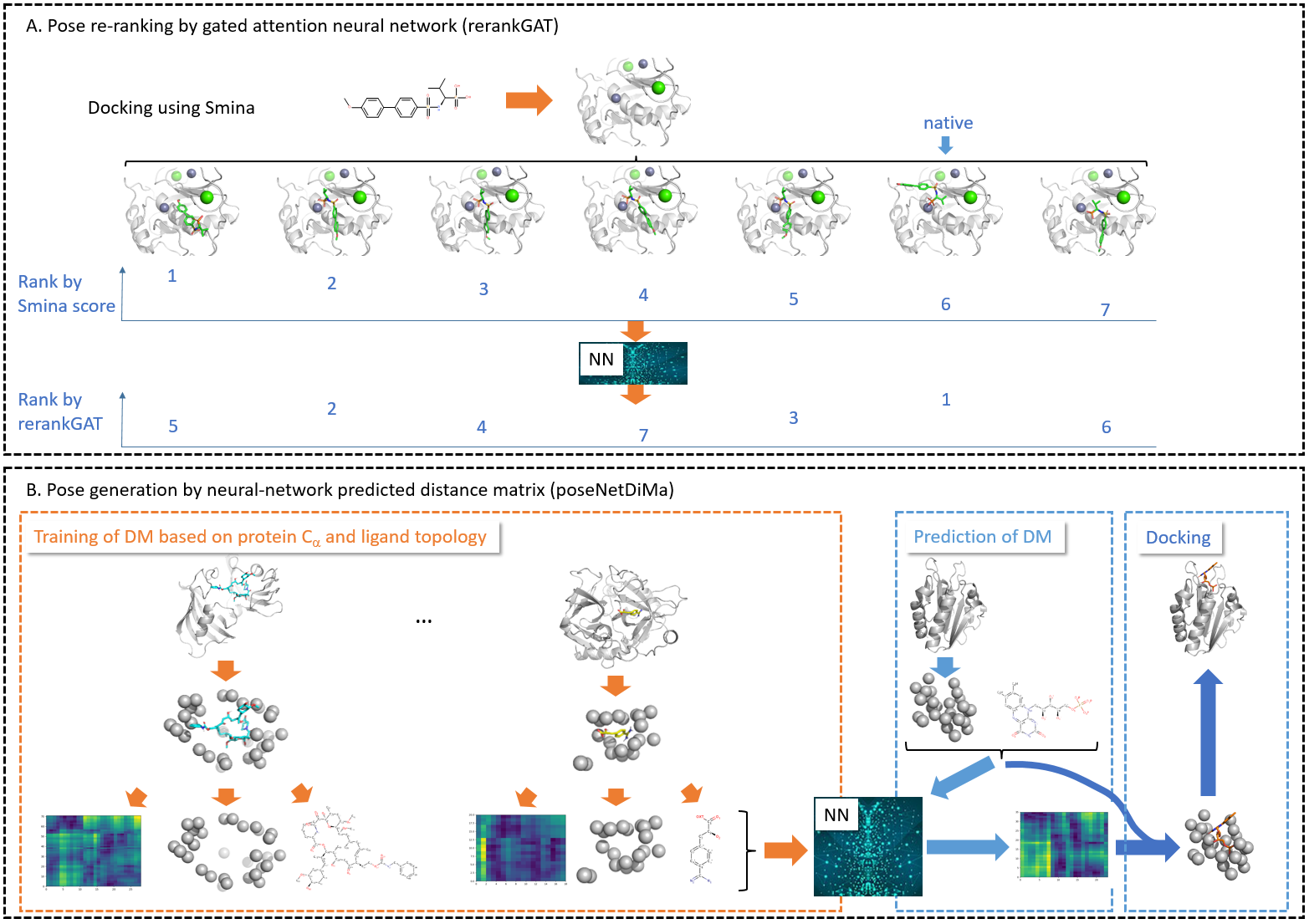}
 \caption{Two neural network approaches to improve flexible docking performance. A. Gated attention neural network (RerankGAT) re-ranks poses obtained from standard docking program, here Smina, aiming to improve pose scoring. B. Graph neural network that predicts distance matrix between ligand atoms and protein C$_\alpha$ atoms (PoseNetDiMa) using protein C$_\alpha$ atom coordinates and ligand topology as input. The predicted distance matrix can be directly used to generate poses with implicit inclusion of side chain flexibility.}
 \label{fig:GeneralSchemeNN}
\end{figure}

In rigid docking applications deep-learning methods proved to be useful in achieving unprecedented accuracy in pose prediction  \cite{mahmoud2020elucidating,ghanbarpour2020fly,Masters2018}. In this work, we aim to utilize deep learning to improve the quality of flexible docking approaches.
We will demonstrate that our deep learning-based concepts increase the accuracy of flexible docking with concurrent improvement in sampling efficiency. Two neural-network-based methods are presented here making initial steps towards this overarching aim (Figure \ref{fig:GeneralSchemeNN}). The first method, named re-ranking by gated attention neural network (RerankGAT) method, re-evaluates docking poses generated by standard docking approaches, while the second method, named pose generation by neural-network predicted distance matrix (PoseNetDiMa), is based on predicting and utilizing the distance matrix, which represents the relation between ligand and protein atoms, for pose generation and ranking.

Concepts using predictions of distance matrices have proven to be useful in many branches of bioinformatics and cheminformatics. AlphaFold, for example,  enabled the an initio prediction of 3D-protein structures with a higher accuracy than any other state-of-the-art methods \cite{senior2020improved,senior2019protein}. Another example used Wasserstein generative adversarial networks (GAN) to generate valid conformations of organic molecules \cite{hoffmann2019generating}.
Our concept of using the prediction of distance matrices between proteins and ligands is the first in the domain of docking or protein-ligand interactions in general.

The RerankGAT method uses a graph representation for each possible ligand-protein pose and a distance-aware gated graph attention mechanism in order to learn to classify the ligand poses\cite{lim2019predicting,ryu2018deeply,sun2019graph}. Details are discussed in the following Materials and Methods section. It is important to emphasize that the RerankGAT methods in this work can boost the ranking of docking poses in case of docking success but cannot address sampling failure, i.e. failure to generate native-like poses independent of their subsequent ranking.

\begin{figure}
 \centering
 
 \includegraphics[width=\textwidth]{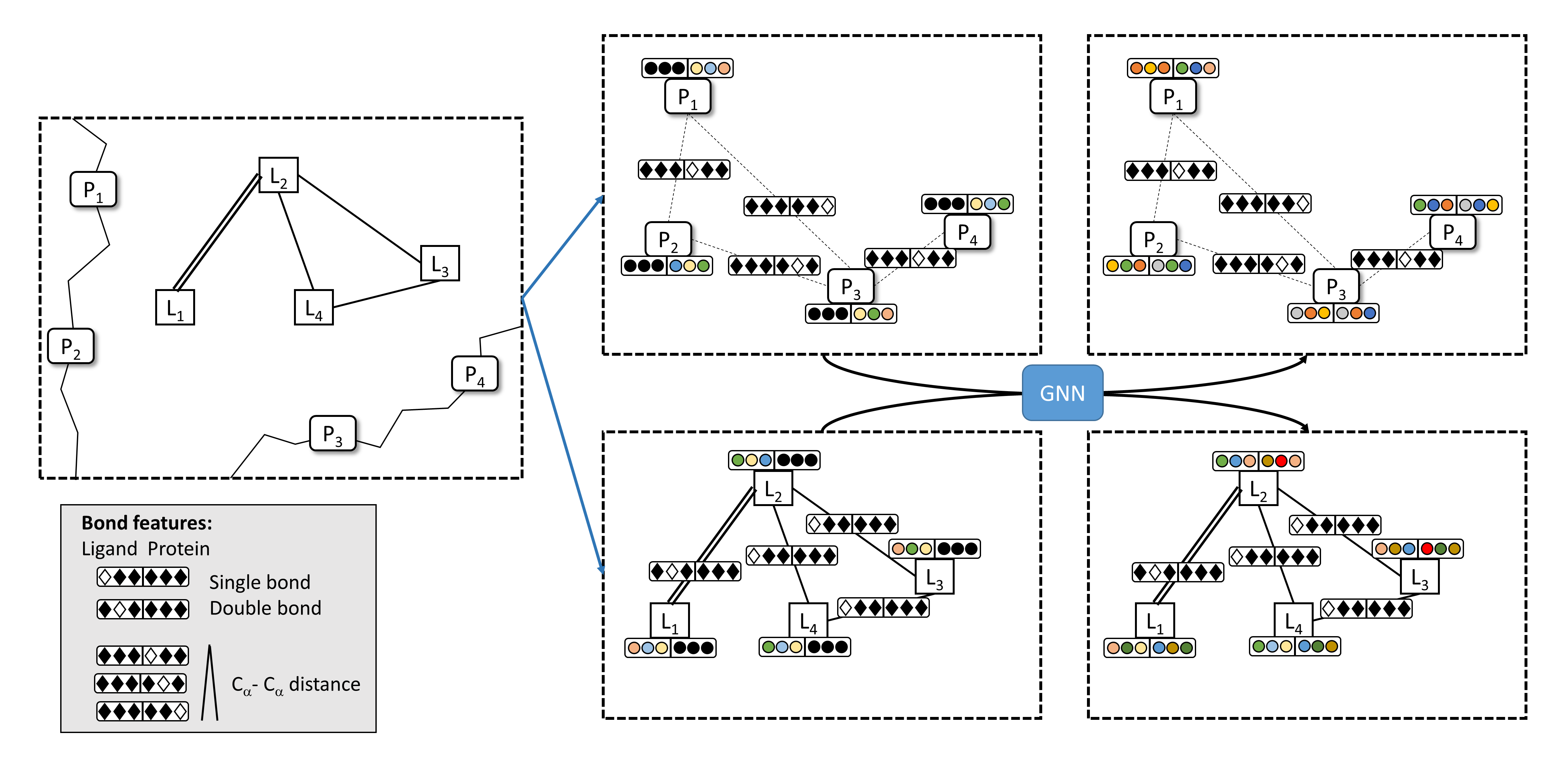}
 \caption{Two different types of graphs are encoded using graph neural network (GNN). One graph used the C$_{\alpha}$ atoms of the binding site residues as nodes colored by type of amino acid. Edges are colored based on distance between connecting C$_{\alpha}$ atoms. The second graph encodes the ligand topology using all heavy atoms. The atom nodes are colored by atom properties, the edges by bond character.}
 \label{fig:SchemeDM1}
\end{figure}

The PoseNetDiMa method predicts the distance matrix between the C$_{\alpha}$ atoms of the target protein and all heavy atoms of the ligand to be docked. C$_{\alpha}$ atoms were chosen to implicitly include side-chain flexibility without explicit sampling. The method uses coordinates of the C$_{\alpha}$ atoms and the ligand topology as input. A graph neural network with a global attention mechanism is trained to predict the pairwise distances between protein C$_{\alpha}$ and ligand heavy atoms\cite{morris2019weisfeiler,lei2017deriving,jin2017predicting,coley2019graph}. The model relies on the heterogeneous graph attention concept \cite{wang2019heterogeneous} where two different types of graphs are encoded (Figure \ref{fig:SchemeDM1}): One graph represents the protein using the C$_{\alpha}$ atoms as nodes colored by the type of amino acid. Edges between nodes are defined based on the Euclidean distance between C$_{\alpha}$ atoms. The other graph encodes the ligand, where nodes and edges are represented by heavy atoms and covalent bonds. No spatial information is included in the ligand graph, as information about ligand conformation has to be predicted in the docking stage based on the predicted distance matrix. In our implementation both graphs are fused into one graph with extended feature vectors as described in subsequent sections. 

The PoseNetDiMa method is far more versatile since the predicted distance matrix can be used in mulitple ways: First, the predicted distance matrix can provide restrains necessary to confine the solutions within a limited space, thus enabling exploration of configurational space in a reasonable time. 
Second, generated poses can be filtered according to their correlation with the predicted distance matrix. 
Third, direct reconstruction of the ligand within the binding pocket based on the distance matrix is possible. 
In summary, the predicted distance matrix can be used in machine-learning assisted docking or re-scoring of poses. Thus, in contrast to RerankGAT, PoseNetDiMa was designed to increase the number of systems for which native docking solutions are identified compared to standard flexible docking. 
 

\section{Materials and Methods}

\subsection{Training Data}\label{Training}
The general set from PDBbind was used for training and initial validation of the models. 
To generate poses for model training and validation, flexible docking was performed using Smina \cite{koes2013lessons}. Unbiased selection of flexible residues was chosen, where any residue was considered flexible if it is located within 4 \si{\angstrom} of the ligand in the X-ray complex structure. The search volume was defined by the centroid of the co-crystallized ligand, adding a padding of 8 \si{\angstrom} to the box encompassing the ligand. Exhaustiveness is set to 8 with 50 modes being requested and using 8 threads per docking job. 
For validation purposes, the data set was split into four groups and 4-fold cross-validation was performed.

\subsection{Validation and Quality Assessment using Cross Docking}

The quality of the models was further assessed using cross docking. The data sets used for this assessment comprises a large number of targets which vary among each other according to their difficulty in flexible docking. The total number of ligands which were docked were around 4500 ligands from 95 targets with an average of 45 ligands per target (Table \ref{tab:cross_docking_set}). All ligands were docked with flexible side chains with the same settings used for docking the General Set from PDBBind (cf. Section \nameref{Training}). The dataset is consistent in coverage with the Disco dataset \cite{wierbowski2020cross}. In contrast to the study of Wierbowski et al. \cite{wierbowski2020cross}
flexible docking on one template protein structure was performed instead of rigid cross-docking.

\begin{longtable}{|p{0.2in}|p{0.6in}|p{0.5in}|p{3.2in}|p{0.4in}|}
\caption{Targets and number of ligands used in cross-docking experiments.}
\label{tab:cross_docking_set}
\\
\hline 
\endfirsthead
\caption{Continued}\\
\hline
\endhead
\endfoot
 & Target & PDB & Description & $N^o$ \\ \hline 
1 & THB & 1Q4X & Thyroid hormone receptor beta-1 & 14 \\ \hline 
2 & WEE1 & 3BIZ & Serine/threonine-protein kinase WEE1 & 8 \\ \hline 
3 & KITH & 2B8T & Thymidine kinase & 2 \\ \hline 
4 & ADRB1 & 2VT4 & Beta-1 adrenergic receptor & 15 \\ \hline 
5 & HDAC2 & 3MAX & Histone deacetylase 2 & 3 \\ \hline 
6 & ANDR & 2AM9 & Androgen Receptor & 92 \\ \hline 
7 & XIAP & 3HL5 & Inhibitor of apoptosis protein 3 & 21 \\ \hline 
8 & MCR & 2AA2 & Mineralocorticoid receptor & 18 \\ \hline 
9 & PRGR & 3KBA & Progesterone receptor & 18 \\ \hline 
10 & GCR & 3BQD & Glucocorticoid receptor & 17 \\ \hline 
11 & ESR2 & 2FSZ & Estrogen receptor beta & 32 \\ \hline 
12 & FPPS & 1ZW5 & Farnesyl diphosphate synthase & 30 \\ \hline 
13 & FA10 & 3KL6 & Coagulation factor X & 104 \\ \hline 
14 & HMDH & 3CCW & HMG-CoA reductase & 19 \\ \hline 
15 & ADRB2 & 3NY8 & Beta-2 adrenergic receptor & 9 \\ \hline 
16 & GLCM & 2V3F & Beta-glucocerebrosidase & 9 \\ \hline 
17 & RXRA & 1MV9 & Retinoid X receptor alpha & 42 \\ \hline 
18 & PA2GA & 1KVO & Phospholipase A2 group IIA & 8 \\ \hline 
19 & HIVRT & 3LAN & Human immunodeficiency virus type 1 reverse transcriptase & 169 \\ \hline 
20 & ESR1 & 1SJ0 & Estrogen receptor alpha & 104 \\ \hline 
21 & GRIA2 & 3KGC & Glutamate receptor ionotropic, AMPA 2 & 84 \\ \hline 
22 & KIF11 & 3CJO & Kinesin-like protein 1 & 32 \\ \hline 
23 & FKB1A & 1J4H & FK506-binding protein 1A & 26 \\ \hline 
24 & VGFR2 & 2P2I & Vascular endothelial growth factor receptor 2 & 24 \\ \hline 
25 & THRB & 1YPE & Thrombin & 216 \\ \hline 
26 & TRY1 & 2AYW & Trypsin I & 169 \\ \hline 
27 & RENI & 3G6Z & Renin & 46 \\ \hline 
28 & PGH1 & 2OYU & Cyclooxygenase-1 & 19 \\ \hline 
29 & TGFR1 & 3HMM & TGF-beta receptor type I & 15 \\ \hline 
30 & PPARA & 2P54 & Peroxisome proliferator-activated receptor alpha & 15 \\ \hline 
31 & JAK2 & 3LPB & Tyrosine-protein kinase JAK2 & 48 \\ \hline 
32 & AKT1 & 3CQW & Serine/threonine-protein kinase AKT & 11 \\ \hline 
33 & PDE5A & 1UDT & Phosphodiesterase 5A & 26 \\ \hline 
34 & MAPK2 & 3M2W & MAP kinase-activated protein kinase 2 & 13 \\ \hline 
35 & LKHA4 & 3CHP & Leukotriene A4 hydrolase & 39 \\ \hline 
36 & HS90A & 1UYG & Heat shock protein HSP 90-alpha & 175 \\ \hline 
37 & CAH2 & 1BCD & Carbonic anhydrase II & 242 \\ \hline 
38 & BRAF & 3D4Q & Serine/threonine-protein kinase B-raf & 46 \\ \hline 
39 & PNPH & 3BGS & Purine nucleoside phosphorylase & 6 \\ \hline 
40 & NRAM & 1B9V & Neuraminidase & 12 \\ \hline 
41 & KIT & 3G0E & Stem cell growth factor receptor & 6 \\ \hline 
42 & HIVPR & 1XL2 & Human immunodeficiency virus type 1 protease & 394 \\ \hline 
43 & UROK & 1SQT & Urokinase-type plasminogen activator & 14 \\ \hline 
44 & HIVINT & 3NF7 & Human immunodeficiency virus type 1 integrase & 8 \\ \hline 
45 & HXK4 & 3F9M & Hexokinase type IV & 24 \\ \hline 
46 & CDK2 & 1H00 & Cyclin-dependent kinase 2 & 310 \\ \hline 
47 & MK10 & 2ZDT & c-Jun N-terminal kinase 3 & 59 \\ \hline 
48 & DEF & 1LRU & Peptide deformylase & 10 \\ \hline 
49 & PGH2 & 3LN1 & Cyclooxygenase-2 & 30 \\ \hline 
50 & PPARD & 2ZNP & Peroxisome proliferator-activated receptor delta & 21 \\ \hline 
51 & MMP13 & 830C & Matrix metalloproteinase 13 & 28 \\ \hline 
52 & MK14 & 2QD9 & MAP kinase p38 alpha & 176 \\ \hline 
53 & PTN1 & 2AZR & Protein-tyrosine phosphatase 1B & 74 \\ \hline 
54 & PUR2 & 1NJS & GAR transformylase & 9 \\ \hline 
55 & MK01 & 2OJG & MAP kinase ERK2 & 69 \\ \hline 
56 & DPP4 & 2I78 & Dipeptidyl peptidase IV & 71 \\ \hline 
57 & DYR & 3NXO & Dihydrofolate reductase & 11 \\ \hline 
58 & ADA & 2E1W & Adenosine deaminase & 15 \\ \hline 
59 & MET & 3LQ8 & Hepatocyte growth factor receptor & 49 \\ \hline 
60 & FAK1 & 3BZ3 & Focal adhesion kinase 1 & 21 \\ \hline 
61 & ROCK1 & 2ETR & Rho-associated protein kinase 1 & 12 \\ \hline 
62 & ACE & 3BKL & Angiotensin-converting enzyme & 8 \\ \hline 
63 & PLK1 & 2OWB & Serine/threonine-protein kinase PLK1 & 10 \\ \hline 
64 & MP2K1 & 3EQH & Dual specificity mitogen-activated protein kinase kinase 1 & 8 \\ \hline 
65 & ACES & 1E66 & Acetylcholinesterase & 41 \\ \hline 
66 & ITAL & 2ICA & Leukocyte adhesion glycoprotein LFA-1 alpha & 13 \\ \hline 
67 & IGF1R & 2OJ9 & Insulin-like growth factor I receptor & 13 \\ \hline 
68 & TRYB1 & 2ZEC & Tryptase beta-1 & 13 \\ \hline 
69 & FA7 & 1W7X & Coagulation factor VII & 42 \\ \hline 
70 & ABL1 & 2HZI & Tyrosine-protein kinase ABL & 39 \\ \hline 
71 & GRIK1 & 1VSO & Glutamate receptor ionotropic kainate 1 & 23 \\ \hline 
72 & ADA17 & 2OI0 & ADAM17 & 12 \\ \hline 
73 & BACE1 & 3L5D & Beta-secretase 1 & 262 \\ \hline 
74 & SRC & 3EL8 & Tyrosine-protein kinase SRC & 48 \\ \hline 
75 & PARP1 & 3L3M & Poly [ADP-ribose] polymerase-1 & 17 \\ \hline 
76 & LCK & 2OF2 & Tyrosine-protein kinase LCK & 32 \\ \hline 
77 & CSF1R & 3KRJ & Macrophage colony stimulating factor receptor & 12 \\ \hline 
78 & CP2C9 & 1R9O & Cytochrome P450 2C9 & 3 \\ \hline 
79 & ALDR & 2HV5 & Aldose reductase & 3 \\ \hline 
80 & SAHH & 1LI4 & Adenosylhomocysteinase & 3 \\ \hline 
81 & AOFB & 1S3B & Monoamine oxidase B & 3 \\ \hline 
82 & PPARG & 2GTK & Peroxisome proliferator-activated receptor gamma & 116 \\ \hline 
83 & FABP4 & 2NNQ & Fatty acid binding protein adipocyte & 22 \\ \hline 
84 & AKT2 & 3D0E & Serine/threonine-protein kinase AKT2 & 10 \\ \hline 
85 & HDAC8 & 3F07 & Histone deacetylase 8 & 7 \\ \hline 
86 & EGFR & 2RGP & Epidermal growth factor receptor erbB1 & 81 \\ \hline 
87 & PYGM & 1C8K & Muscle glycogen phosphorylase & 18 \\ \hline 
88 & CXCR4 & 3ODU & C-X-C chemokine receptor type 4 & 4 \\ \hline 
89 & PYRD & 1D3G & Dihydroorotate dehydrogenase & 4 \\ \hline 
90 & AMPC & 1L2S & Beta-lactamase & 59 \\ \hline 
91 & FNTA & 3E37 & Protein farnesyltransferase / geranylgeranyltransferase type I alpha subunit & 23 \\ \hline 
92 & CASP3 & 2CNK & Caspase-3 & 21 \\ \hline 
93 & TYSY & 1SYN & Thymidylate synthase & 11 \\ \hline 
94 & AA2AR & 3EML & Adenosine A2a receptor & 2 \\ \hline 
95 & CP3A4 & 3NXU & Cytochrome P450 3A4 & 7 \\ \hline 
\end{longtable}

\subsection{Model Features}

For both network models, basic chemical properties of atoms were used as initial node features. Those features include an atom's elemental type, connectivity index, aromaticity, implicit valence,  partial charge estimates,  number of attached hydrogen atoms, surface area contributions (Labute ASA in rdkit and TPSA), Crippen LogP, Crippen MR and electro-topological State descriptors known as EState in rdkit. Bond featurization depends on the bond type, bond conjugation and whether the bond is a ring bond.
In the PoseNetDiMa model, however, which relies on a coarse-grained representation of the protein, the nodes in the protein graph represent the C$_{\alpha}$ atoms of the amino acids of the binding site with one hot-encoding for the twenty different amino acids. Features describing the physicochemical properties of each amino acid (i.e. polar, charged, hydrophobic, aromatic side chain) are added to the feature vector describing the nodes of the protein graph. 
To generate a single, heterogeneous graph containing both ligand and protein nodes, the ligand and protein node features are concatenated into one feature vector  $F_i = (\f_i^{(L)},\f_i^{(P)})$.
Since C$_{\alpha}$ atoms are not covalently bonded, the edges were represented using virtual bonds that reflect pairwise distances between two nodes being within 7 \si{\angstrom}. In detail, five distance bins between 2 \si{\angstrom} and 7 \si{\angstrom} are generated, and an edge between two C$_{\alpha}$ atoms within a maximum distance of 7 \si{\angstrom} is colored by its association with the matching distance bin using one-hot encoding. For example, an edge between two C$_{\alpha}$ atoms with distance of 5.7 \si{\angstrom} will have the feature vector (0, 0, 0, 1, 0). To combine the heterogeneous bond features of ligand and protein, the feature vectors are concatenated: $F_{ij} = (\f_{ij}i^{(L)},\f_{ij}^{(P)})$.

\subsection{Graph Neural Network Models}

\begin{figure}[h]
 \centering
 \includegraphics[width=\textwidth]{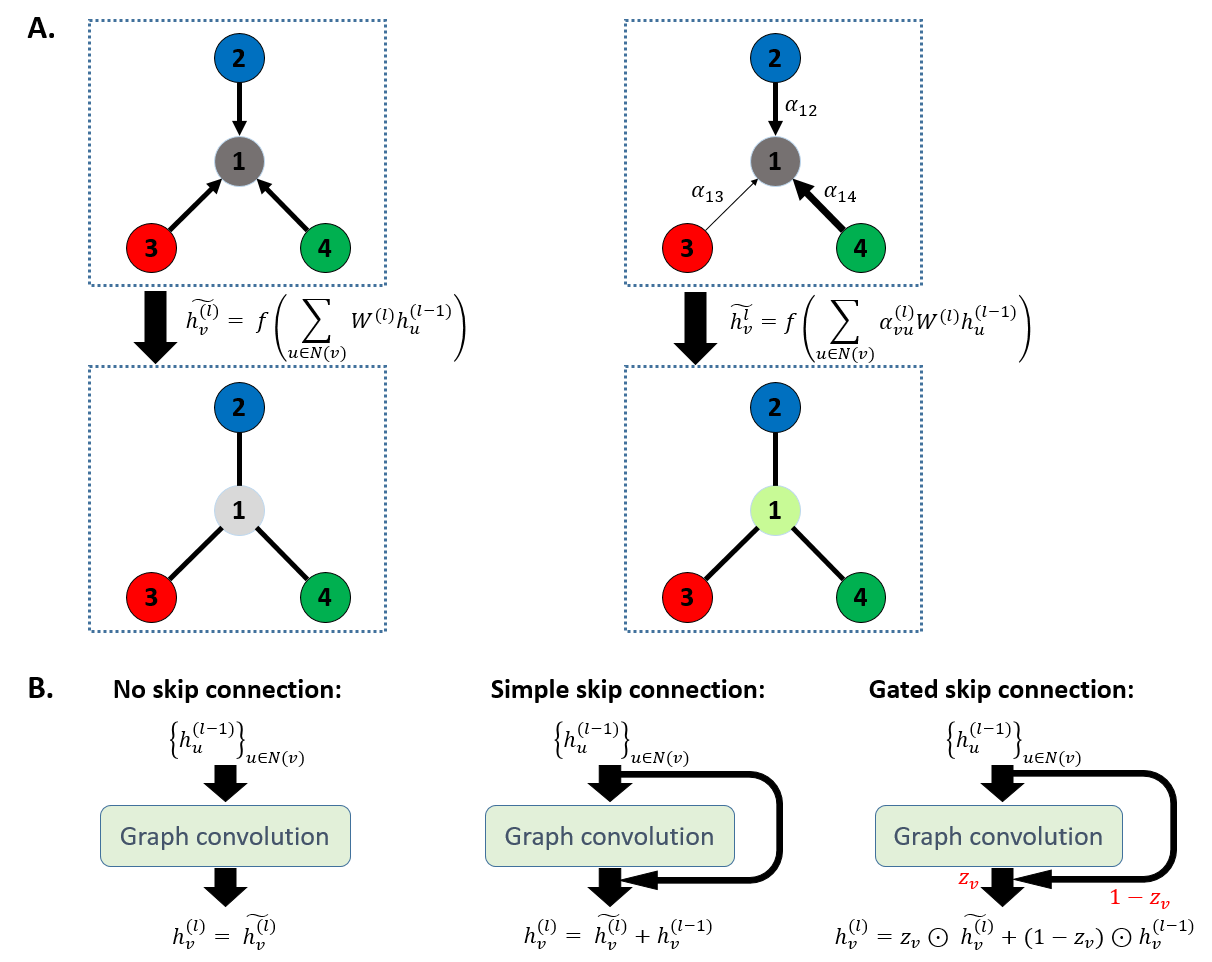}
 \caption{A. Scheme of a graph convolution step (\textit{left}) and its attention-augmented version (\textit{right}).
 Central nodes update is carried out using neighboring nodes where different width of arrows reflect the importance of information transfer, hence attention. B. Different versions of skip connections to conserve initial node features over mulitple update steps. Skip rate $z_v$ is determined using neural network layer. }
 \label{fig:GAT}
\end{figure}

\subsubsection{Pose re-ranking model: RerankGAT}
The model for pose re-ranking utilizes the Graph Neural Network algorithm  described by Lim et al. \cite{lim2019predicting} The network was gated with an attention mechanism that takes distances into consideration. 

\subsubsection{Model for prediction of distance matrix: PoseNetDiMa}
The PoseNetDiMa model is inspired by the work of Jin et al. \cite{jin2017predicting,coley2019graph} on synthesis prediction. In our work, a similar network as in Jin et al. \cite{jin2017predicting,coley2019graph} is used with the main aim to predict protein-ligand distance matrices that could be used as distance restraints during pose sampling or for pose filtering and re-scoring. The network tries to identify the correspondence between each atom of the ligand and C$_{\alpha}$ atoms comprising the protein-binding site .  

\begin{figure}[h]
 \centering
 \includegraphics[width=\textwidth]{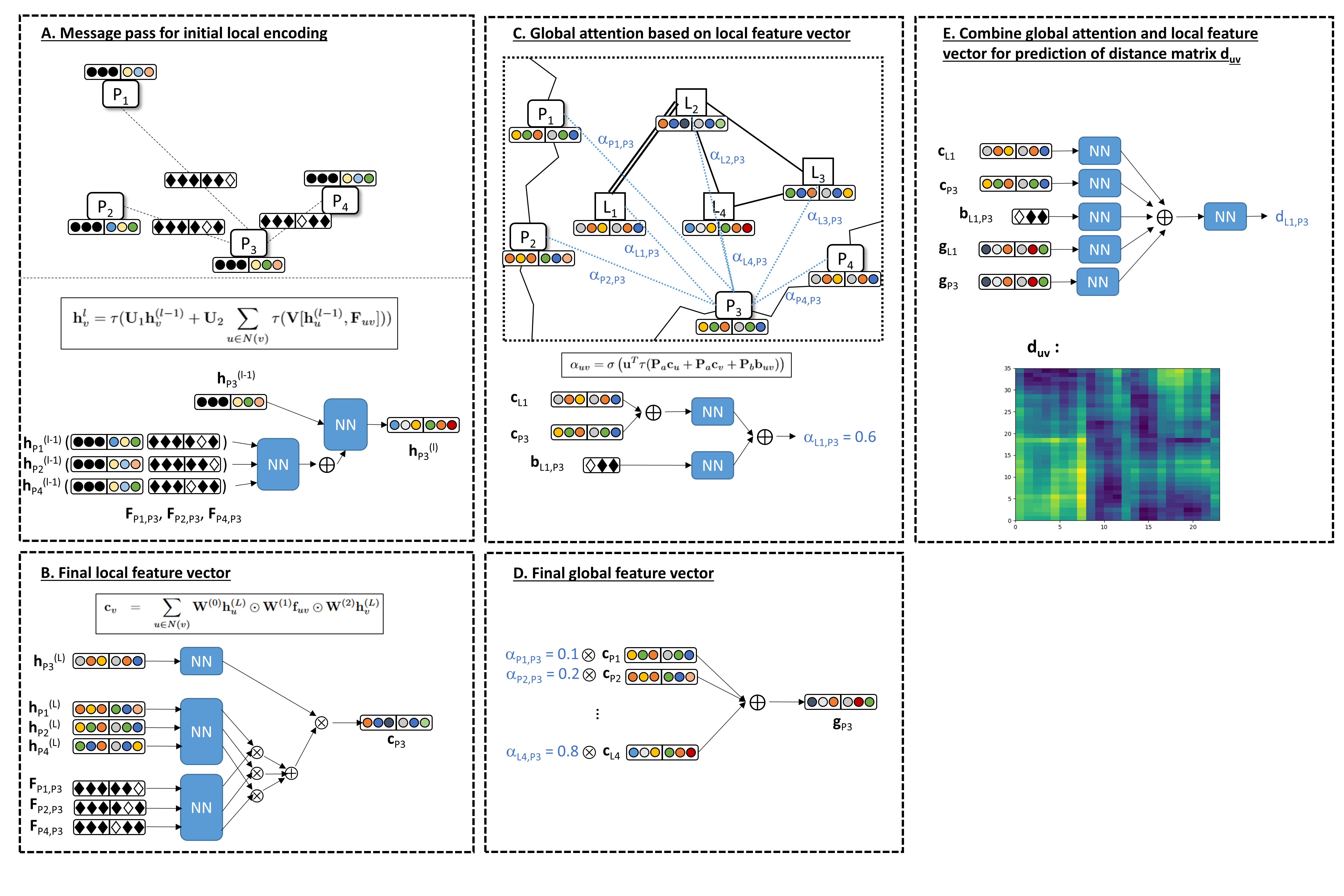}
 \caption{Scheme of PoseNetDiMa to predict distance matrix based on coarse grained representation of protein and 2D representation of ligand. After initial local encoding using message pass (A), a local feature vector is determined based on combining atom encoding and bond featurization for protein and ligand separately (B). Using global attention (C) protein and ligand encodings are combined and a final global feature vector is computed (D). Local and global feature vector are finally combined to predict the protein-ligand distance matrix (E).}
 \label{fig:DM2}
\end{figure}

In a first step, the nodes of both protein and ligand graph are encoded using a graph neural network (Figure \ref{fig:DM2} A). 
To encode the hidden features $\mbf{h}_v^{(l)}$ of a node $v$ in layer $l$, messages $\mbf{m}_{uv}$ from neighboring nodes $u \in N(v)$ are collected. 
To compute the message between $u$ and $v$, the current node feature $\mbf{h}_u^{(l-1)}$ and edge feature $\mbf{F}_{uv}$ are concatenated and used as input of a neural network (Figure \ref{fig:DM2} A) with ReLU activation function $\tau$:
\begin{equation}
\mbf{m}_{vu} = \tau(\mbf{V} [\mbf{h}_u^{(l-1)},\mbf{F}_{uv}])
\end{equation}
After collecting all message from neighboring nodes, the previous hidden feature $\mbf{h}_v^{(l-1)}$ of node $v$ is added in a skip connection and a subsequent neural network finally encodes new hidden features

\begin{equation}
\mbf{h}_v^{(l)} = \tau(\mbf{U}_1 \mbf{h}_v^{(l-1)} + \mbf{U}_2\sum_{u\in N(v)} \tau(\mbf{V} [\mbf{h}_u^{(l-1)},\mbf{F}_{uv}]))
\end{equation}
where $\mbf{h}_v^{(0)}=\mbf{F}_v$ and $\mbf{U}_1,\mbf{U}_2,\mbf{V}$ are shared weights. 

After $L$ steps of graph convolutions, the current hidden feature vector $\mbf{h}_v^{(l)}$ is transformed into a final local feature vector $\mbf{c}_v$ (Figure \ref{fig:DM2} B). First, $\mbf{h}_v^{(l)}$, neighboring feature vectors $\mbf{h}_u^{(l)}$ and corresponding edges undergo additional tensor multiplications by $\W^{(0)}$, $\W^{(1)}$ and $\W^{(2)}$. Subsequent Hadamard products between the three resulting entities generates the final local feature vector $\mbf{c}_v$:
\begin{eqnarray}
\cc_v &=& \W^{(2)} \h_v^{(L)} \odot  \sum_{u\in N(v)} \W^{(0)} \h_u^{(L)} \odot \W^{(1)}\mbf{F}_{uv}  \label{eq:nn-edge} 
\end{eqnarray}
$\cc_v$ is a feature vector  that locally encodes the chemical environment of the atom.

To predict the likely distance between ligand atoms and protein $C_\alpha$ atoms, the currently distinct ligand and protein graphs need to be connected. In other words, information needs to be shared between both graphs. 
The main idea is that the interaction strengths between different residue types and ligand atom types varies (e.g. hydrogen bonds differ in distance dependency and strength compared to hydrophobic contacts). 
Whereas the local environment of a node within one of the graph is captured according to each atom's connectivity with its neighbors, the global environment, i.e. protein-ligand interactions, is incorporated  through a global attention mechanism which allows for weighted information exchange between nodes of the two different graphs (Figure \ref{fig:DM2} C). 
The attention score $\alpha_{uv}$ between nodes $u$ and $v$ is derived by
\begin{eqnarray}
\alpha_{uv} = \sigma\left(\mbf{u}^T \tau(\mbf{P}_a \cc_u + \mbf{P}_a \cc_v + \mbf{P}_b \mbf{b}_{uv})\right)
\end{eqnarray}
where $\sigma(\cdot)$ is the sigmoid function, and $\mbf{b}_{uv}$ is a feature vector that represents information about the relationship between $u$ and $v$, i.e. whether the two nodes represent protein-protein, ligand-protein or ligand-ligand pairwise interactions or covalent bonds.

The global feature representation $g_u$ of node $u$ is then calculated as the weighted sum of all other surrounding nodes where the weights correspond to the attention factors (Figure \ref{fig:DM2} D): 
\begin{eqnarray}
\mbf{g}_u &=& \sum_v \alpha_{uv} \cc_v
\end{eqnarray} 

Finally, the distance between two nodes $u$ and $v$, e.g. ligand atom $u$ and protein C$_\alpha$ atom $v$ is computed (Figure \ref{fig:DM2} E) by
\begin{eqnarray}
d_{uv} &=& \mbf{u}^T \tau(\mbf{M}_a \mbf{g}_u + \mbf{M}_a \mbf{g}_v + \mbf{M}_b \mbf{f}_{uv} + \mbf{P}_a \cc_u + \mbf{P}_a \cc_v)
\end{eqnarray}

{\bf Training.} The network is finally trained to reproduce the experimentally measured distances $y_{uv}$ between ligand atoms $u$ and protein C$_\alpha$ atoms $v$. 
The in-dependant prediction of each label is performed due to the quadratic complexity of the problem. The interaction labels can be determined by the product of  $N$ ligand atoms and $M$ C$_{\alpha}$ atoms and this quadratic complexity prevents higher-order predictions.  

\subsubsection{Docking using PoseNetDiMa}
\begin{figure}[h]
 \centering
 \includegraphics[width=\textwidth]{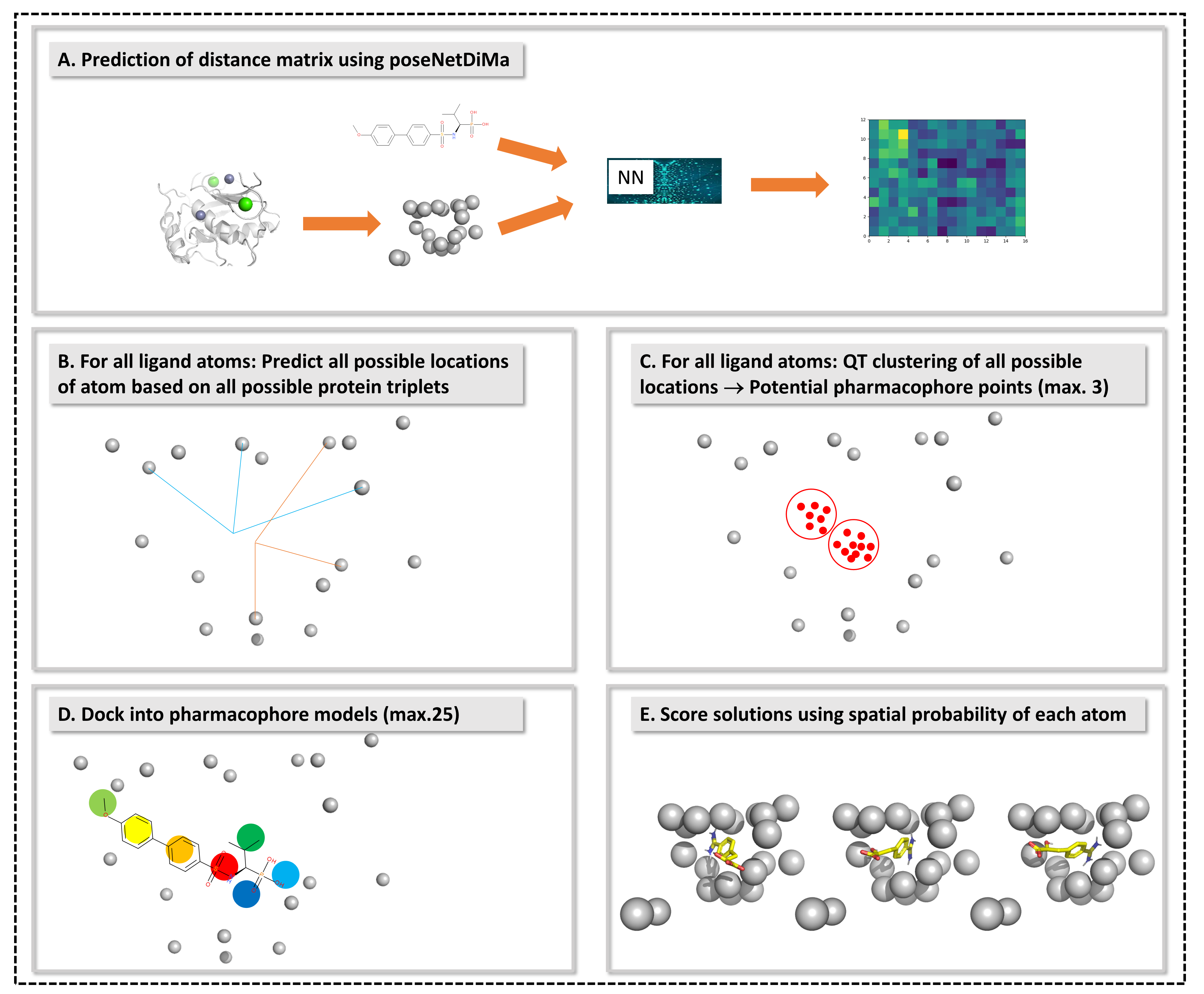}
 \caption{Scheme for using PoseNetDiMa for docking. A. Distance matrix is predicted using PostNetDiMa. B. For every ligand atom, all possible locations are computed based on the predicted distance matrix using all possible protein triplets. C. Those points are clustered using QT clustering algorithm. D. A maximum of 25 pharmacophore models are generated by random selection of combinations of cluster centers. Docking is performed to pharmacophore models using LSAlign. E. Those poses are rescored using iDock and atomic density maps obtained from predicted distance matrix.}
 \label{fig:PoseNetDiMa_Docking}
\end{figure}

Similar to RerankGAT, PoseNetDiMa can be used for re-ranking poses obtained from standard flexible docking such as Smina. 
As Smina is unable to generate native-like poses for a large number of targets, we tested if the predicted distance matrices could be directly used for both posing and ranking phases in docking.
A scheme of the overall docking scheme based on PostNetDiMa is shown in Figure \ref{fig:PoseNetDiMa_Docking}.

First, the distance matrix for the protein-ligand system of interest is predicted using PostNetDiMa. For every ligand atom, all possible locations are computed based on the predicted distance matrix using all possible triplets of C$_\alpha$ atoms in the binding site.  Those points are clustered using Quality Threshold (QT) clustering algorithm with a radius of 1 \AA. Clustering is stopped either when half of all possible points are assigned to cluster or a maximum number of three clusters is identified for an atom. Pharmacophore models are generated from the clusters with one element per atom. For each atom the used cluster center is selected randomly, generating a maximum number of 25 pharmacophores. Docking is performed to the pharmacophore models using LSAlign \cite{hu2018ls}. Those poses are rescored using iDock on atomic density maps. Those density maps are 3D grids where the density of a ligand atom $i$ at grid point $k$ is obtained from the product of  normal distribution functions centered around the predicted distance $d_{ji}$ between ligand atom $i$ and C$_\alpha$ atom $j$ 
\begin{equation}
    p_k^i = \prod_j \exp{(-0.5 \cdot (r_{jk} - d_{ji})^2)}
\end{equation}
where $r_{jk}$ is the distance between protein atoms $j$ and grid point $k$.

\section{Results}

\subsection{General Set Evaluation}
\begin{figure}[h]
 \centering
 \includegraphics[width=0.75\textwidth]{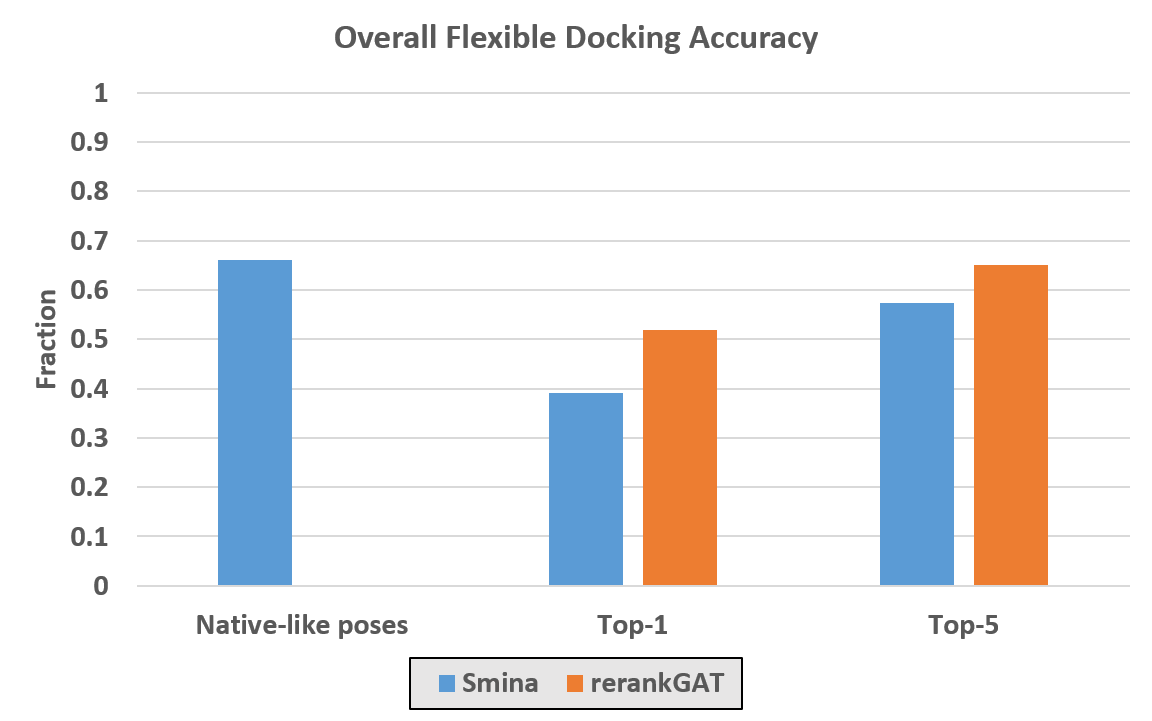}
 \includegraphics[width=0.75\textwidth]{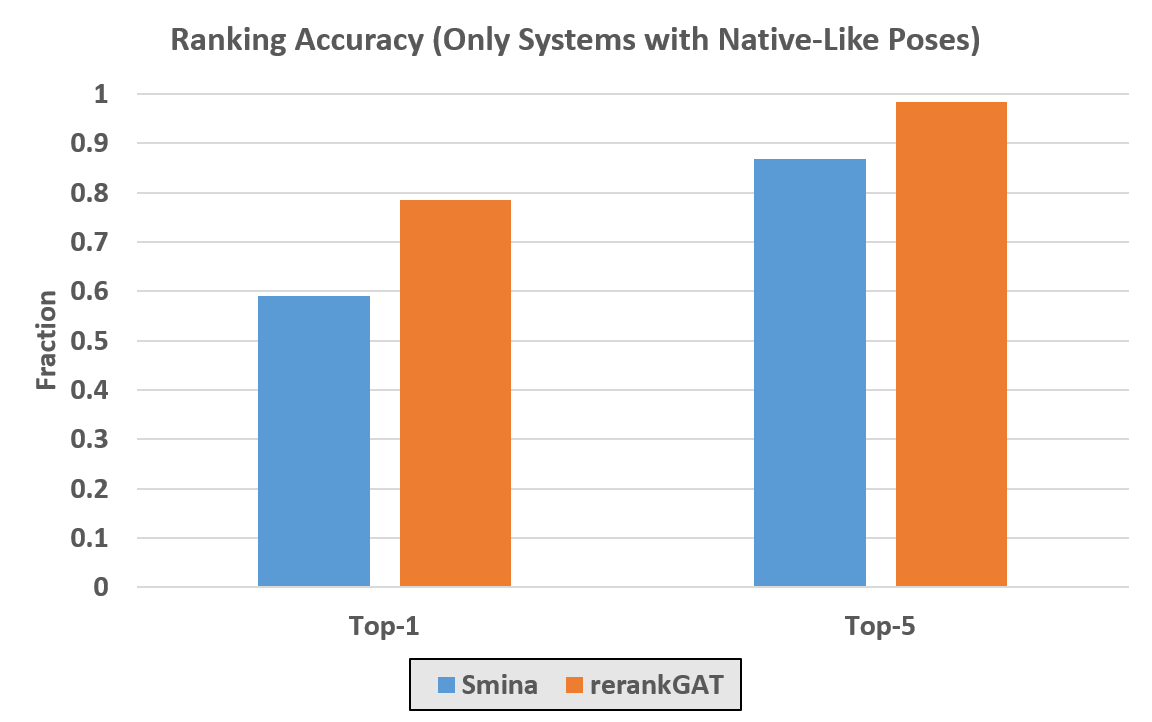}
 \caption{Ranking performance using Smina and rerankGAT (top) on all systems and (bottom) on systems with at least one native-like pose.}
 \label{fig:AccuracyRerankGAT}
\end{figure}

Four-fold cross validation was carried out using the General-set from PDBbind. The cumulative results of only the test sets in the four cross-validation runs are reported. 
First, we tested the re-ranking performance of RerankGAT based on the poses obtained from Smina docking.
Smina was only able to generate native-like poses (RMSD $<$ 2 \AA\ to native pose) for 66 \% of all systems (Figure \ref{fig:AccuracyRerankGAT}, top). 
For those systems with native-like poses, Smina ranks 59 \% of them as top-1 and 87 \% within the top-5 poses (Figure \ref{fig:AccuracyRerankGAT}, bottom).

Figure \ref{fig:AccuracyRerankGAT} shows that the RerankGAT deep-learning model could considerably boost the ranking of pose-prediction in case a native-like pose was sampled by Smina.
For almost all systems, for which a native-like pose was sampled, that pose was retrieved within the top-5 re-ranked poses. 
For 78 \% of systems with native-like pose, that pose was ranked at top-1 position.

\begin{figure}[h]
 \centering
 \includegraphics[width=0.75\textwidth]{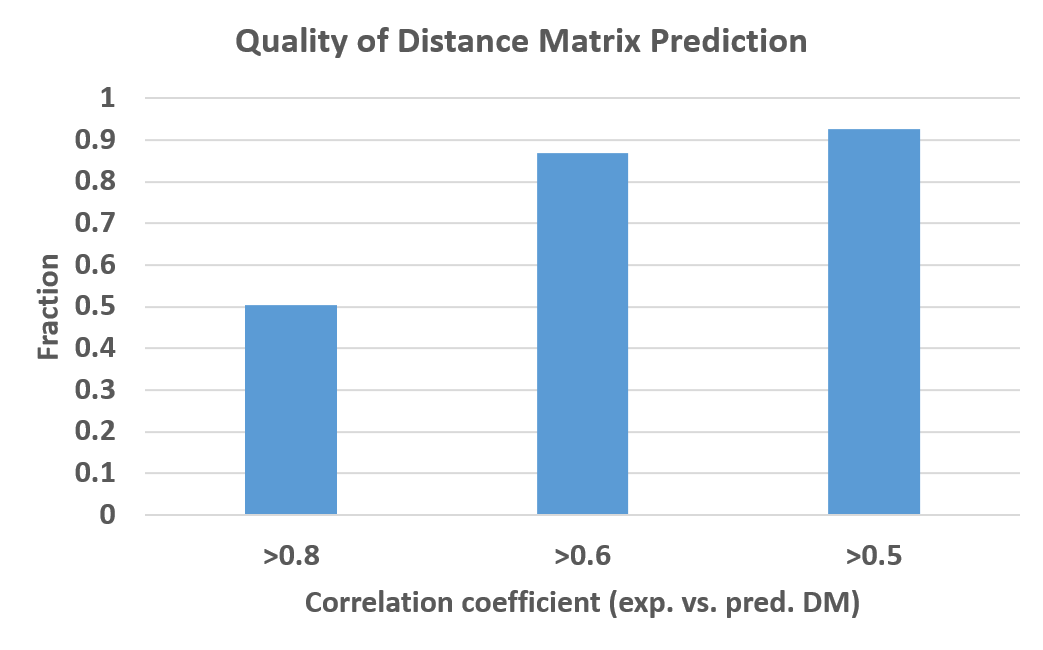}
 \includegraphics[width=0.75\textwidth]{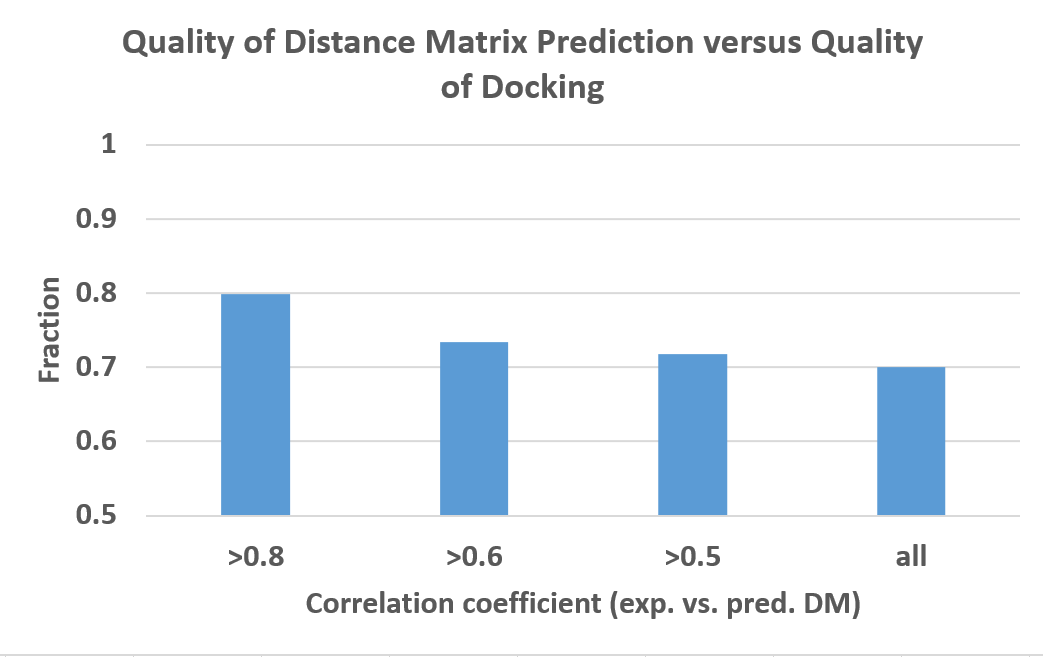}
 \caption{(Top) Fraction of systems with certain distance matrix prediction accuracy measured by correlation coefficient between experimental and predicted distance matrix. (Bottom) Fraction of systems with native-like pose using Smina correlated with the distance matrix prediction accuracy.}
 \label{fig:AccuracyDM}
\end{figure}

In response to the high number of systems, for which no native-like pose could be generated using Smina, PoseNetDiMa was designed to generate poses based on the predicted distance matrix between protein residues and ligand atoms. 
First, we investigated the quality of PoseNetDiMa to predict experimental distance matrices. Figure \ref{fig:AccuracyDM} (top) shows that for the majority of systems a correlation with $r^2>0.5$ could be achieved, for half of the systems a correlation even larger than 0.8.
Interestingly, there is a  correlation between number of systems with native-like poses and quality in distance matrix prediction (Figure \ref{fig:AccuracyDM}, bottom).
For example, 80 \% of systems with high distance-matrix quality ($r^2>0.8$) have near-native poses, while only 50 \% with poor distance matrix quality ($r^2<0.5)$.
Initial analysis indicates that the distance matrix for systems with high flexibility and particular solvent-exposure, that may have alternative binding poses, are difficult to predict. Those systems also show no robust prediction in binding poses in docking \cite{majewski2019protein}.

\subsection{Cross-Docking Assessment}
\begin{figure}[h]
 \centering
 \includegraphics[width=0.75\textwidth]{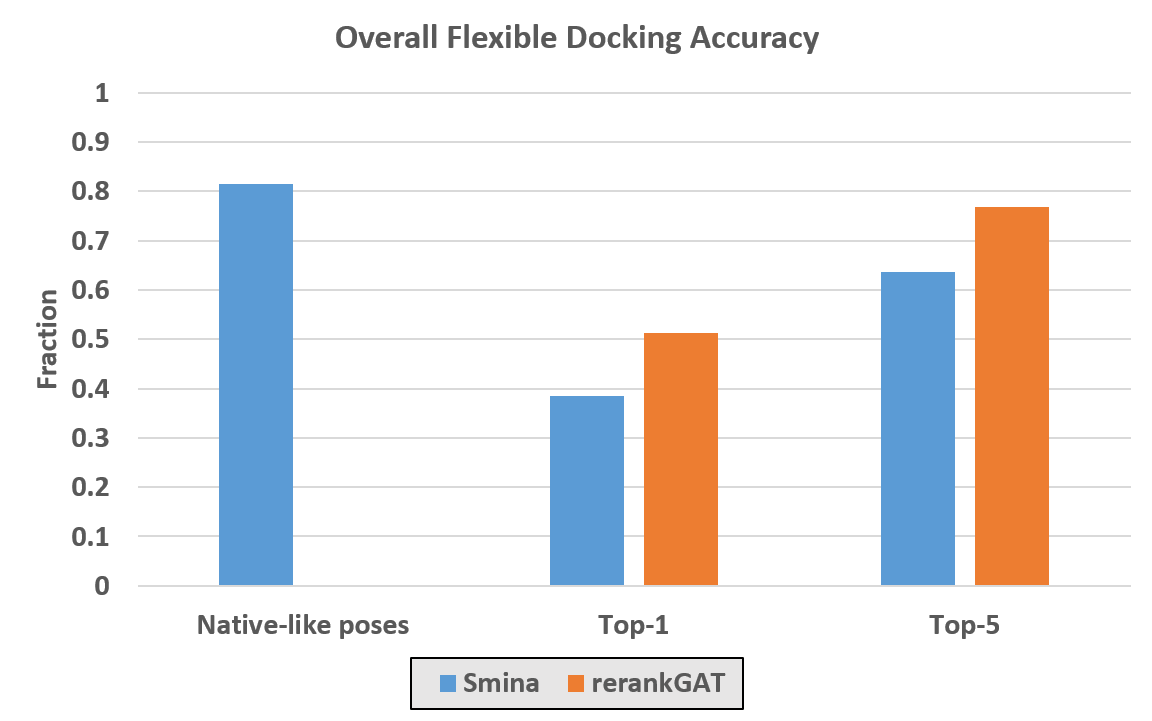}
 \includegraphics[width=0.75\textwidth]{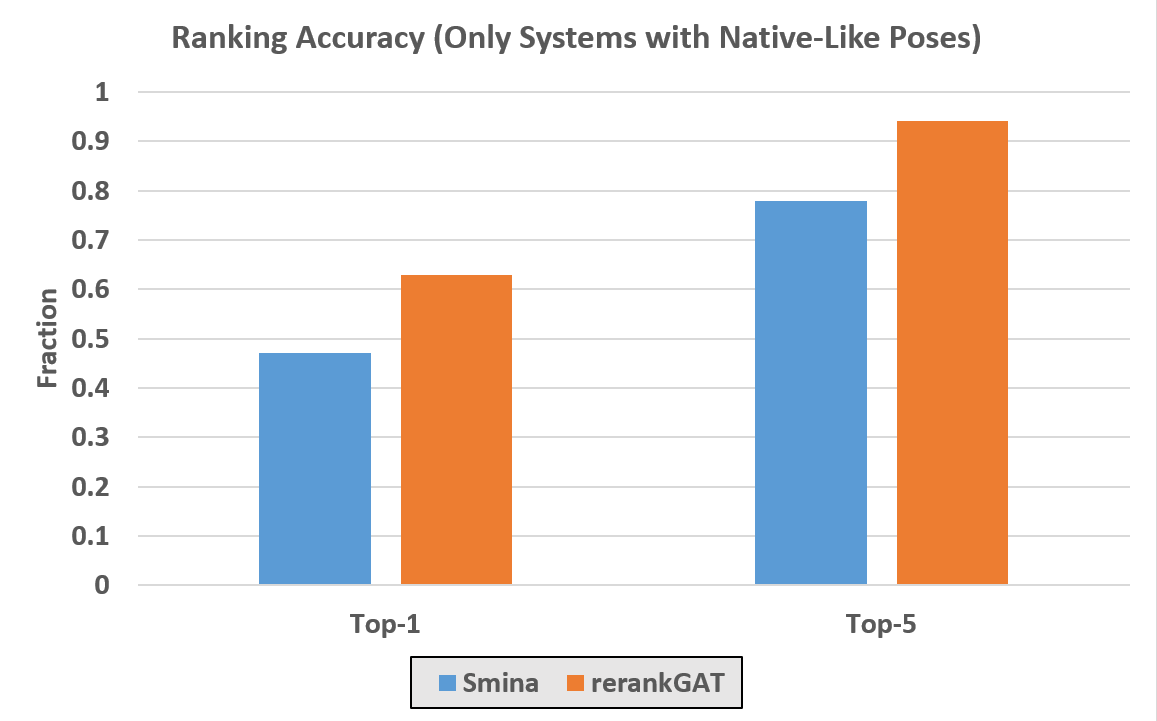}
 \caption{Ranking performance using Smina and rerankGAT (top) on all systems and (bottom) on systems with at least one native-like pose for cross-docking data set.}
 \label{fig:AccuracyRerankGATCross}
\end{figure}

For additional validation, the same analysis was performed on cross docking on 95 targets with different levels of difficulty. Some targets are known to have high failure rate in cross docking such as Cytochrome P450 3A4 and Caspase-3. 
Smina was used for flexible docking of the cross-docking dataset and the poses were re-scored using the graph-attention neural network model trained on the general set of PDBbind (Figure \ref{fig:AccuracyRerankGATCross}).
The first observation is that there are a higher number of systems with native-like poses compared to the general set of PDBbind (81 \% versus 66 \%).
The reason for this difference is that for each target system, the protein structure with the highest success rate was selected for  cross-docking studies following a previous protocol \cite{wierbowski2020cross}.

The higher number of systems with native-like poses resulted in higher number of native-like poses identified in the top-5 ranked list with rescoreGAT (77 \%) again outperforming Smina scoring (64 \%).
In contrast, the performance for identifying a native-like pose in the top-1 position remained unchanged in the cross-docking study compared to flexible docking to PDBbind.

\begin{figure}[h]
 \centering
 \includegraphics[width=0.75\textwidth]{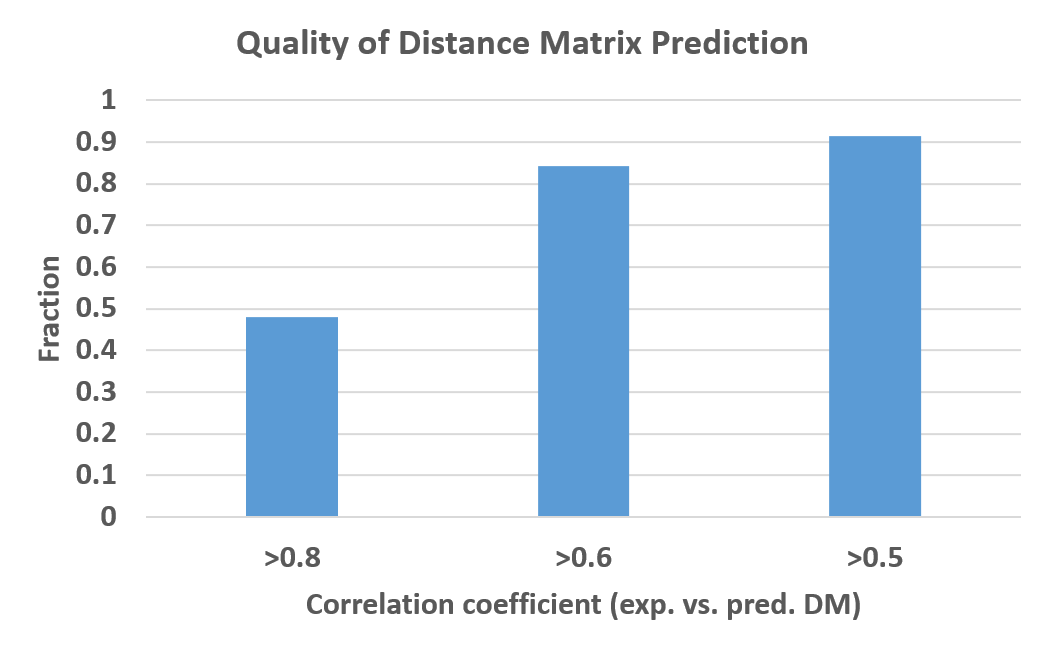}
 \includegraphics[width=0.75\textwidth]{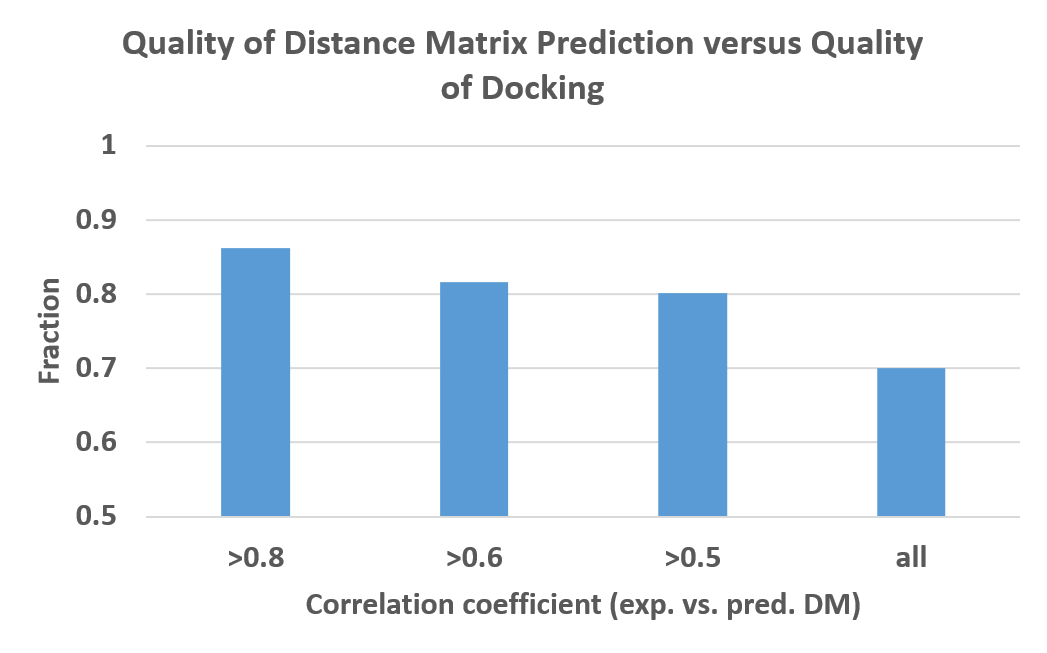}
 \caption{(Top) Fraction of systems with certain distance matrix prediction accuracy measured by correlation coefficient between experimental and predicted distance matrix. (Bottom) Fraction of systems with native-like pose using Smina correlated with the distance matrix prediction accuracy. Data for cross-docking data set is shown.}
 \label{fig:AccuracyDMCross}
\end{figure}

Figure \ref{fig:AccuracyDMCross} shows similar accuracy in predicting the distance matrix for the cross-docking dataset compared to the PDBbind dataset, and the same trend of overall better prediction of the distance matrix for systems with more likely success in generating native-like poses.

\subsubsection{Re-ranking of poses using PoseNetDiMa}
\begin{figure}[h]
 \centering
 \includegraphics[width=\textwidth]{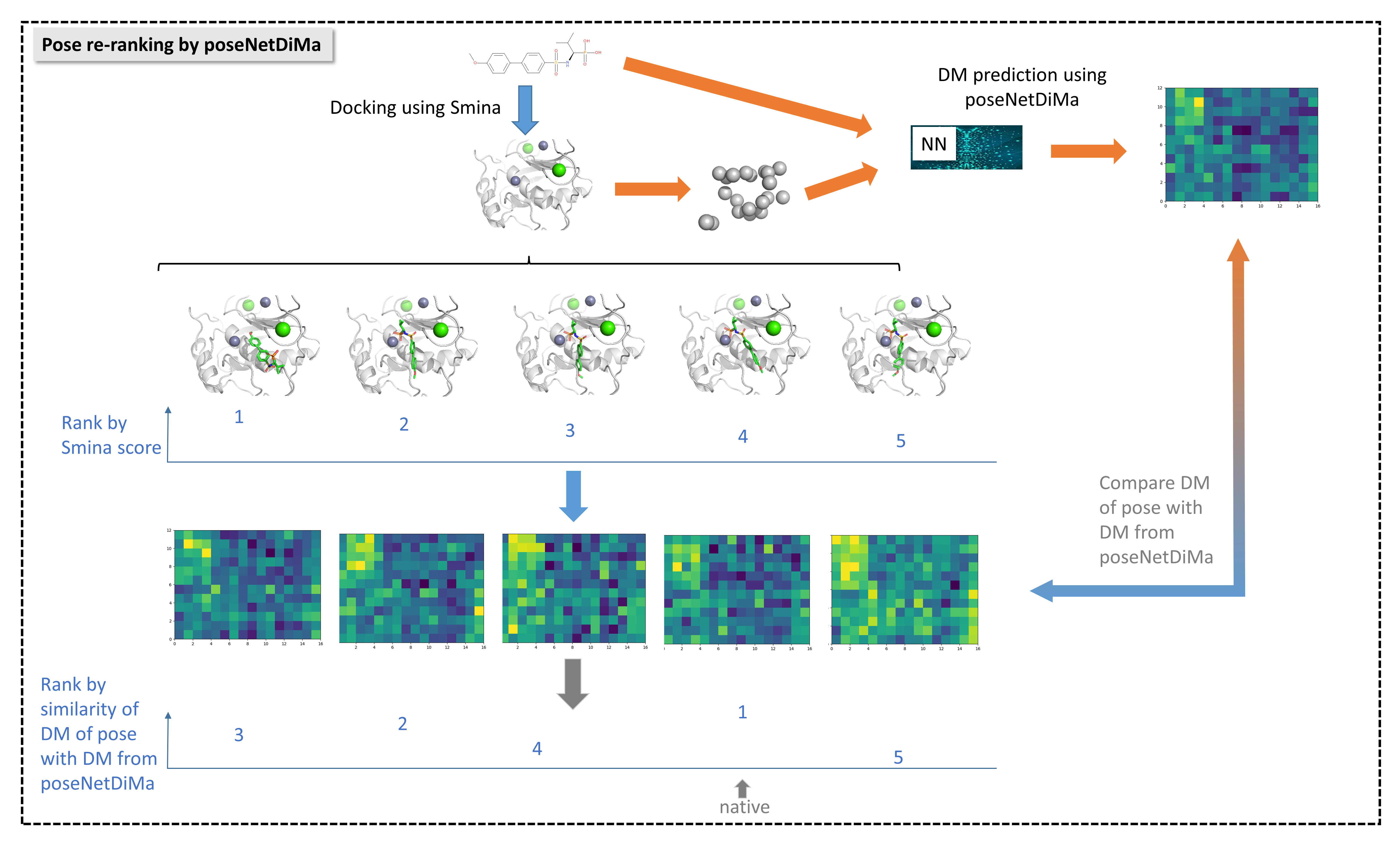}
 \caption{Scheme for pose re-ranking using poseNetDiMa. Protein-ligand distance matrix is predicted using postNetDiMa and compared with corresponding distance matrices measured for each docking pose. Re-ranking is performed based on similarity between predicted distance matrix and distance matrix of a given docking pose.}
 \label{fig:RerankingPoseNetDiMa}
\end{figure}

Next, we explored the potential of PoseNetDiMa to re-rank poses obtained from Smina.
Whereas the similarity between native and docked pose is typically measured by their RMSD value, alternatively the similarity of their corresponding protein-ligand distance matrices could be used (Figure \ref{fig:RerankingPoseNetDiMa}). 
Thus, assuming the distance matrix predicted by PoseNetDiMa is similar to the experimentally known distance matrix, the docked poses could be translated into distance matrices and ranked by their similarity to the predicted distance matrix,
Based on this idea, the hypothesis has been that pose ranking could be improved using the predicted distance matrix from PoseNetDiMa.

\begin{figure}[h]
 \centering
 \includegraphics[width=0.49\textwidth]{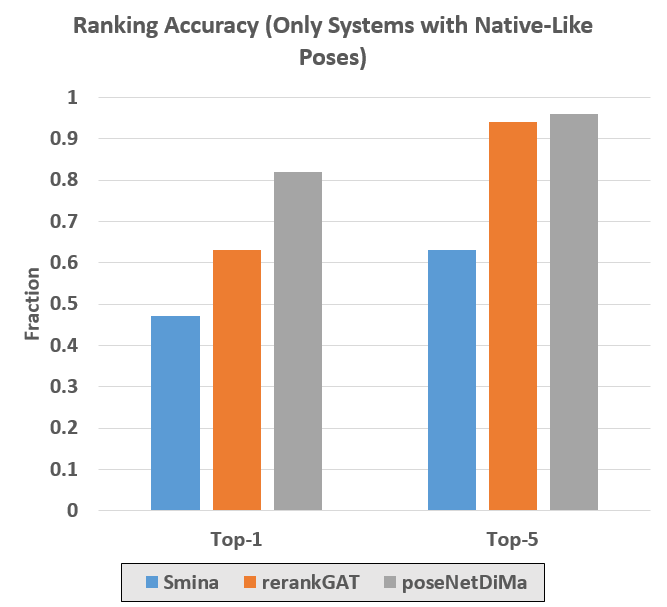}
 \includegraphics[width=0.49\textwidth]{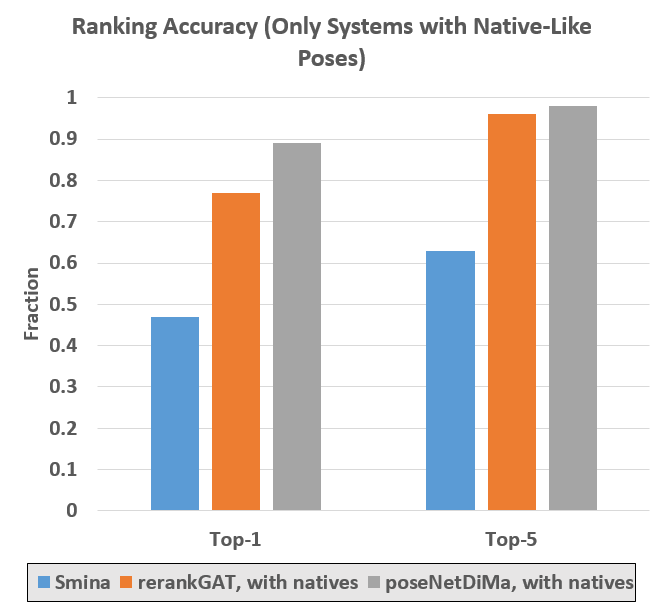}
 \caption{Re-ranking accuracy of docking poses obtained from Smina using PoseNetDiMa for systems with native-like poses using only docked poses (left) or when adding native poses from X-ray structure (right).}
 \label{fig:AccuracyRerankingPoseNetDiMa}
\end{figure}

The analysis was performed on those systems for which the docking engine was able to generate near-native poses. 
As shown in Figure \ref{fig:AccuracyRerankingPoseNetDiMa} (left), PoseNetDiMa significantly improves pose ranking, even outperforming RerankGAT by a significant margin. 
Whereas, Smina is only able to rank 47 \% of systems with native-like poses as top-1, PoseNetDiMa increases this percentage to 82 \%. 
Adding native poses to the pool of docked poses further increases this percentage to 89 \%. 

\subsubsection{Docking using PoseNetDiMa}
\begin{figure}[h]
 \centering
 \includegraphics[width=0.49\textwidth]{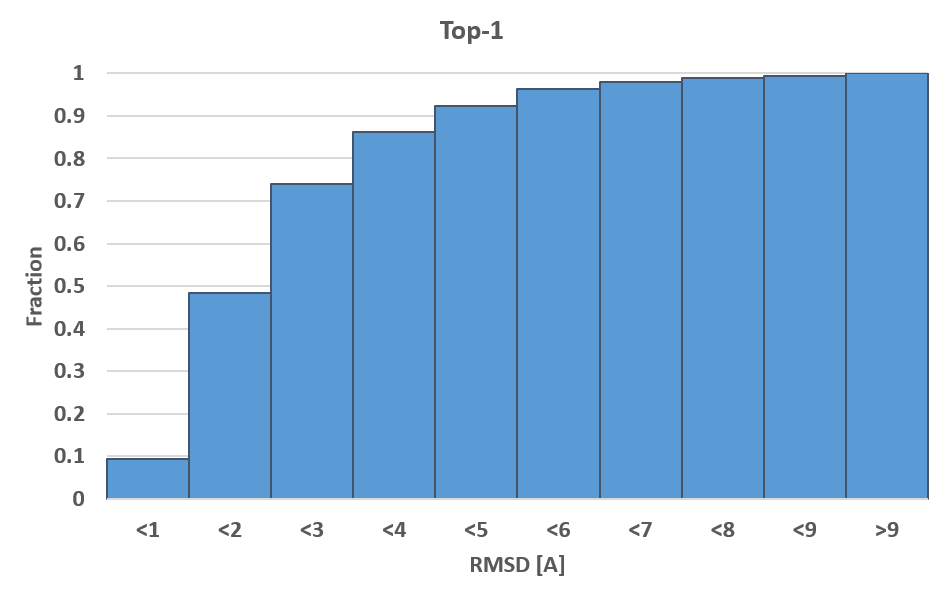}
 \includegraphics[width=0.49\textwidth]{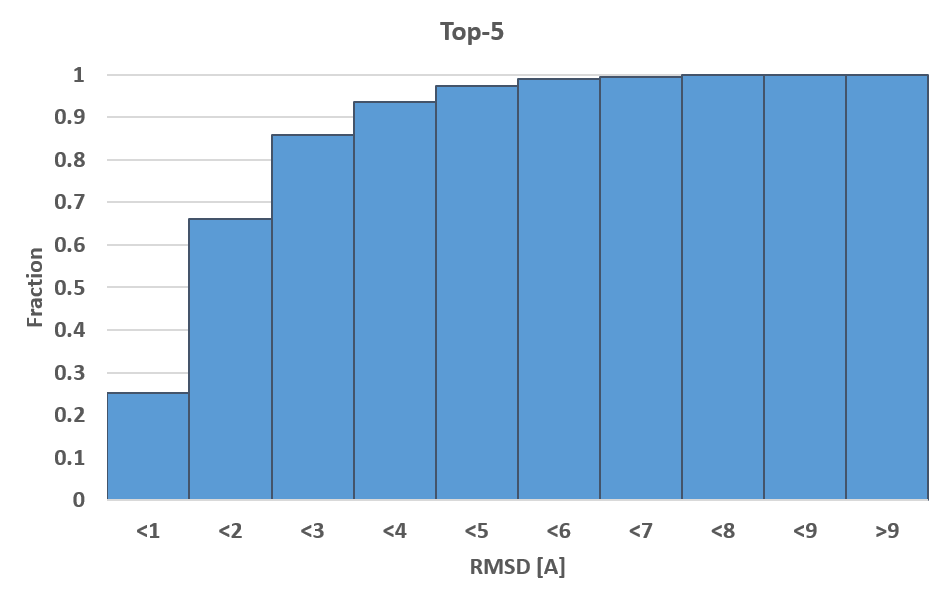}
 \caption{Cumulative probability of predicting docking pose within certain RMSD to native binding mode at top-1 or among top-5 ranked solutions using PostNetDiMa in docking modus.}
 \label{fig:DockingPoseNetDiMa}
\end{figure}

Docking using PoseNetDiMa was performed on the cross-docking set. Despite only using the C$_\alpha$ atoms from the protein, PostNetDiMa obtained the same success rate to identify a native-like pose at top-1 position and even slightly outperformed flexible docking using Smina when considering the top-5 poses (Figure \ref{fig:DockingPoseNetDiMa}). 
Interestingly for almost all systems a top-5 ranked pose was identifed within an RMSD of less than 4 \AA. This means that the general orientation of the scaffold of a ligand could be identified for almost all systems based on a coarse grained representation of the protein.

\begin{figure}[h]
 \centering
 \includegraphics[width=0.49\textwidth]{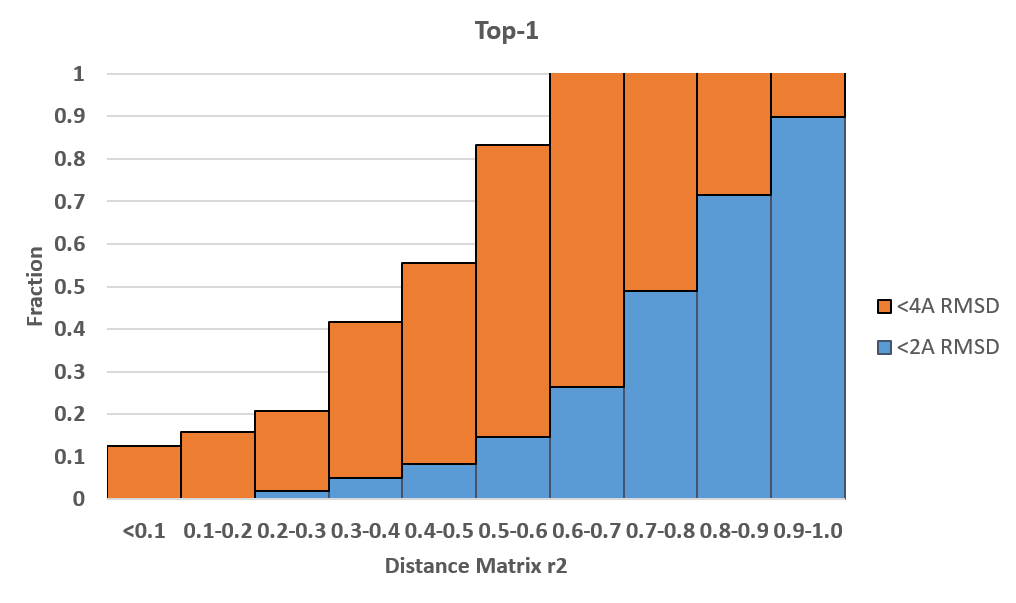}
 \includegraphics[width=0.49\textwidth]{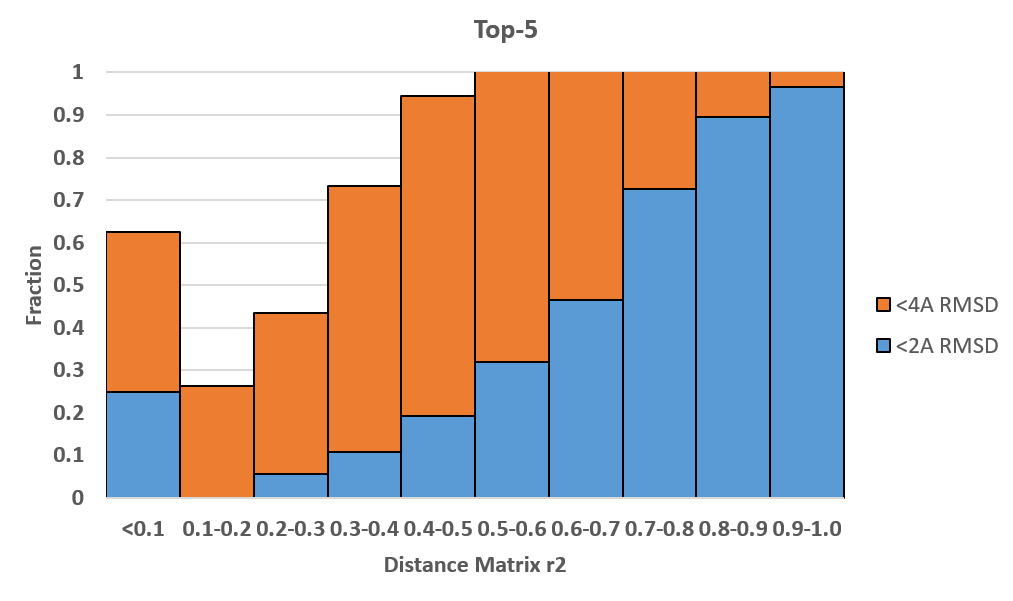}
 \caption{Probability of predicting native-like docking poses within an RMSD of less than 2 \AA (blue) and 4 \AA (orange) to the native binding mode at top-1 or among top-5 ranked solutions using PostNetDiMa in docking modus. Dependency on prediction quality of distance matrix is shown.}
 \label{fig:DockingPoseNetDiMa_r2}
\end{figure}

Furthermore Figure \ref{fig:DockingPoseNetDiMa_r2} highlights a strong correlation between prediction quality of the protein-ligand distance matrix and the docking quality.
In particular, a native-like pose could most likely be generated among the top-5 ranked poses if the correlation coefficient $r^2$ between experimental and predicted distance matrix is larger than 0.8, and such a native-like pose is top ranked if $r^2$ is even larger than 0.9. 
Thus, in the future we will focus on improving the model for predicting the protein-ligand distance matrix, as this will directly improve docking performance beyond the quality of full-atomistic flexible docking programs.

\section{Conclusion}
We demonstrated in this study how flexible docking performance can be significantly improved using deep learning approaches.
Two different models have been designed for this task: RerankGAT, a model based on graph convolutional neural networks, which was used to re-rank existing poses. Besides standard docking algorithms, those poses could also be obtained from molecular dynamics simulations or similarity-based alignment algorithms. 
The second model, PoseNetDiMa, that generates distance matrices between ligands and proteins based on ligand topology and C$_\alpha$ atoms of binding site residues, can also been used for reranking poses. Furthermore, PoseNetDiMa also provides the necessary information to directly guide ligand placement.

Analysis of targets used in flexible docking reveals that standard docking strategies show weak accuracy in binding pose generation for flexible proteins and proteins with large binding sites compared to ligand size, Using distance matrices that are based on C$_\alpha$ atoms only, explicit side-chain sampling becomes obsolete, reducing the degrees-of-freedom significantly. This can result in more efficient and accurate sampling of native-like poses.












\bibliography{acs-achemso}

\end{document}